\documentclass[preprint,showpacs,preprintnumbers,amsmath,amssymb,floats,nofootinbib]{revtex4}
\usepackage[dvips]{graphicx}
\usepackage[english]{babel}
\usepackage{amsmath}
\usepackage{amssymb}
\usepackage{bm}

\newcommand{\be}{\begin{eqnarray}}
\newcommand{\ee}{\end{eqnarray}}
\newcommand{\p}{\partial}
\newcommand{\bp}{\bar{\partial}}

\begin{document}

\title{Topological Soliton with Nonzero Hopf Invariant in Yang-Mills-Higgs Model}
\author{Yan He$^1$ and Hao Guo$^2$}
\affiliation{$^1$College of Physical Science and Technology,
Sichuan University, Chengdu, Sichuan 610064, China}
\affiliation{$^2$Department of Physics, Southeast University, Nanjing 211189, China}
\date{\today}

\begin{abstract}
We propose a topological soliton or instanton solution with nonzero Hopf invariant to the 3+1D non-Abelian gauge theory coupled with scalar fields. This solution, which we call Hopf soliton, represents a spacetime event that makes a $2\pi$ rotation of the monopole. Although the action of this Hopf soliton is logarithmically divergent, it may still give relevant contributions in a finite-sized system. Since the Chern-Simons term for the unbroken $U(1)$ gauge field may appear in the low energy effective theory, the Hopf soliton may possibly generate fractional statistics for the monopoles.
\end{abstract}

\maketitle

\section{Introduction}

Magnetic monopole, although not discovered in Nature, has attracted lots of attentions from the theoretical physics community\cite{Shnir}. The first monopole solution was proposed by Dirac \cite{Dirac}. He argued that the existence of the monopole implies quantized electrical charges because of the minimal coupling of the $U(1)$ gauge field in quantum mechanics. However, in his monopole model, the vector potential is singular at some points located on the Dirac's string. This drawback was later overcome in the t'Hooft-Polyakov magnetic monopole solution \cite{tHooft}, where the $U(1)$ field is embedded in a non-Abelian gauge field coupled with a scalar field. The Derrick theorem \cite{Derrick} implies that there is no stable soliton solution to the scalar field when the spatial dimension is larger than one. However, the coupling between the scalar field and the non-Abelian gauge field helps to stabilize this monopole solution.

The magnetic charge of the t'Hooft-Polyakov monopole can be identified as the topological charge, hence it is naturally connected to the classification of the homotopy group $\pi_2(S^2)$. On the contrary, one may simply think that the magnetic charge of the Dirac's monopole has no topological origin. While it has been shown that the nontrivial $U(1)$ bundle in this model can actually be thought of as a Hopf fabrication \cite{Nakahara}. Consequently, the magnetic charge of the Dirac's monopole is connected to the Hopf invariant of another homotopy group $\pi_3(S^2)$ \cite{Ryder}. Therefore the same charge can be attributed to different homotopy groups because of different gauge field realizations of a monopole.

In general, different homotopy groups, such as $\pi_2(S^2)$ and $\pi_3(S^2)$, give rise to different topological solitons. A common example is the nonlinear sigma model \cite{Belavin}. In the 2+1D case, this model has a skyrmion-type solution \cite{Skyrme} classified by the winding number associated with the group $\pi_2(S^2)$. In a similar way, there is also an instanton-type solution to the same model \cite{Zee} classified by the nonzero Hopf invariant of the group $\pi_3(S^2)$. This instanton solution represents a 2+1D spacetime event that makes a $2\pi$ rotation of the skyrmion. If the Hamiltonian further contains a Chern-Simons term or Hopf term, this instanton can generate fractional statistics for the associated skyrmion.

In this paper, we propose a topological soliton or instanton solution with nonzero Hopf invariant (Hopf soliton) to the 3+1D non-Abelian scalar gauge theory. This solution can be thought of as either a soliton-type solution in the spatial part of the 4+1D theory or an instanton-type solution of the 3+1D theory. The relation between the Hopf soliton and the non-Abelian magnetic monopole is very similar to that between the skyrmion and the instanton in the nonlinear sigma model discussed above. In the 3+1D case, the solution represents a spacetime event which makes a $2\pi$ rotation of the monopole. In the presence of a Chern-Simons term for the unbroken $U(1)$ gauge field, the Hopf soliton may possibly generate fractional statistics for the monopole. This seems to be contradict to the fact that in 3+1D only fermionic and bosonic statistics are possible for point like particles. However, we will later see that the action of Hopf soliton is logarithmically divergent, thus this type of solution can contribute only in a finite-sized system. Because of this reason, one can not neglect the size of the monopole core and treat it as a point-like particle.

The paper is organized as follows. In section \ref{hopfmap}, we give a brief review on the Hopf mapping and Hopf invariant. In section \ref{solu}, we find out the soliton solution with nonzero Hopf invariant. In section \ref{topo}, we discuss the topological and physical meaning of the solution. The conclusion is outlined in section \ref{conclu}.

\section{Hopf mapping and Hopf invariant}
\label{hopfmap}
In this section, we briefly review the Hopf mapping and Hopf invariant, which also helps to identify the topological charge of the Hopf soliton in our later discussions. It is well known that the mapping between two $n$-dimensional spheres is classified by the $n$-th homotopy group $\pi_n(S^n)=\mathcal{Z}$. The geometric meaning of the winding number $k\in \mathcal{Z}$ is that when the pre-image point sweeps around the whole sphere, the image point sweeps the whole sphere $k$ times. While the Hopf mapping is a map between $S^3$ and $S^2$, i.e., spheres with different dimensions. Hence it is classified by the homotopy group $\pi_3(S^2)$. The topologically nontrivial Hopf mapping is characterized by the Hopf invariant $\mathcal{H}$. The geometric meaning of the Hopf invariant is not as intuitive as that of the winding number.

Here we give a simple visualization of the Hopf mapping.
Under a Hopf mapping, the pre-image of a point on $S^2$ is a circle in $S^3$. Hence the pre-images of two different points are two different circles. Under a topologically trivial Hopf mapping, these two circles are not linked. While under a topologically nontrivial Hopf mapping, these two circles are linked together for one time and form a so called Hopf link.

Mathematically, we introduce a pair of complex numbers $z_1=x_1+ix_2$ and $z_2=x_3+ix_4$ to describe $\mathcal{R}^4$, hence the sphere $S^3$ can be characterized by $|z_1|^2+|z_2|^2=1$. The Hopf mapping $f:$ $S^3\to S^2$ is given by $y^a=\bar{z}_i\,\sigma^a_{ij}z_j$ for $a=1,2,3$, where $\sigma^a$ are Pauli matrices. More explicitly, the Hopf mapping is written as
\be \label{Hopf1}
y^1=2(x_1x_3+x_2x_4),\quad  y^2=2(x_1x_4-x_2x_3), \quad y^3=x_1^2+x_2^2-x_3^2-x_4^2.
\ee
and one can verify that $y^ay^a=|\bar{z}_iz_i|^2=1$, then $y^a$ does describe a point on $S^2$.

The Hopf invariant of the above Hopf mapping (\ref{Hopf1}) can be directly evaluated. Let $\Omega_2$ be the volume 2-form of $S^2$. Since $S^2$ is a two dimensional space, then $\Omega_2$ must be closed, thus we trivially have $d\Omega_2=0$. Moreover, $\Omega_2$ must not be exact, otherwise we will have $\Omega=d\alpha$ which implies $\int_{S^2}\Omega_2=\int_{\p S^2}\alpha=0$ by Stokes theorem. This contradicts the fact the volume of $S^2$ is not zero. The Hopf mapping pulls back the volume 2-form $\Omega_2$ from $S^2$ to $S^3$. We define $\omega_2=f^*\Omega_2$ which is again closed. Since the cohomology of $S^3$ is trivial, i.e., $H^2(S^3)=0$, then there is no nontrivial 2-form on $S^3$. Therefore $\omega_2$ must be exact and can be further written as $\omega_2=d\omega_1$ where $\omega_1$ is a 1-form on $S^3$. Finally the Hopf invariant is defined as
\be
\mathcal{H}=\frac{1}{16\pi^2}\int_{S^3}\omega_1\wedge\omega_2.
\ee
It easy to verify that $\mathcal{H}$ is invariant under a continuous deformation of the map.

The evaluation of $\mathcal{H}$ can be conveniently performed by using the cartesian coordinates. The volume 2-form of a unit 2-sphere is given by
\be \label{O2}
\Omega_2=y_1\,dy_2\wedge dy_3-y_2\,dy_1\wedge dy_3+y_3\,dy_1\wedge dy_2.
\ee
Inserting the Hopf mapping (\ref{Hopf1}), after some algebra we find that the pulled back 2-form is given by
\be
\omega_2&=&4\Big[(x_3^2+x_4^2)dx_1\wedge dx_2
+(x_1x_4-x_2x_3)(dx_1\wedge dx_3+dx_2\wedge dx_4)\nonumber\\
& &-(x_1x_3+x_2x_4)(dx_1\wedge dx_4-dx_2\wedge dx_3)+(x_1^2+x_2^2)dx_3\wedge dx_4\Big].
\ee
This expression can be further simplified by noticing $\sum_ix_i^2=1$ and $\sum_ix_idx_i=0$ sucessively. One can verify that
\be
0&=&(\sum_ix_idx_i)(x_1dx_2-x_2dx_1+x_3dx_4-x_4dx_3)\nonumber\\
&=&\Big[(x_1^2+x_2^2)dx_1\wedge dx_2
-(x_1x_4-x_2x_3)(dx_1\wedge dx_3+dx_2\wedge dx_4)\nonumber\\
& &+(x_1x_3+x_2x_4)(dx_1\wedge dx_4-dx_2\wedge dx_3)+(x_3^2+x_4^2)dx_3\wedge dx_4\Big].
\ee
Adding the above two equations together, we find that $\omega_2$ can be rewritten as
\be
\omega_2=4(dx_1\wedge dx_2+dx_3\wedge dx_4).
\ee
Then it is easy to find that
\be
\omega_1=2(x_1dx_2-x_2dx_1+x_3dx_4-x_4dx_3).
\ee
Finally the outer product of the above two differential forms gives the volume element of a unit $S^3$
\be
\omega_1\wedge\omega_2&=&8(x_1\,dx_2\wedge dx_3\wedge dx_4
-x_2\,dx_1\wedge dx_3\wedge dx_4\nonumber\\
& &\quad+x_3\,dx_1\wedge dx_2\wedge dx_4-x_4\,dx_1\wedge dx_2\wedge dx_3).
\ee
Since the surface area of the unit $S^3$ is $2\pi^2$, then the Hopf invariant of the map (\ref{Hopf1}) is
\be
\mathcal{H}=\frac{1}{16\pi^2}\int_{S^3}\omega_1\wedge\omega_2=1.
\ee

For later use, the above statements can be elaborated in a more physical language. We can express the pulled back 2-form $\omega_2$ as a $U(1)$ gauge field strength. Hence $\omega_1$ becomes the corresponding gauge potential. Introducing a set of coordinate parameters $u_{1,2,3}$ to describe $S^3$, then the 2-form $\omega_2$ can again be evaluated by pulling back $\Omega$ given by (\ref{O2})
\be
\omega_2=\frac12F_{\mu\nu}du_{\mu}\wedge du_{\nu},\quad \omega_1=A_{\mu}du_{\mu} \quad \mbox{for } \mu,\nu=1,2,3,
\ee
where in the component form $F_{\mu\nu}$ is the surface area element of $S^2$
\be
F_{\mu\nu}=\epsilon^{ijk}y_i\p_{\mu}y_j\p_{\nu}y_k
=-2i(\p_{\mu}\bar{z}_i\p_{\nu}z_i-\p_{\nu}\bar{z}_i\p_{\mu}z_i).
\ee
In the second equality, we have inserted the Hopf mapping and used the equality (\ref{p2}). We will visit this equality later in details. Moreover,  the gauge potential $A_{\mu}$ can be easily found by $F_{\mu\nu}=\p_{\mu}A_{\nu}-\p_{\nu}A_{\mu}$
\be
A_{\mu}=-i\Big[\bar{z}_i(\p_{\mu}z_i)-(\p_{\mu}\bar{z}_i)z_i\Big].
\ee
In this language, the Hopf invariant can be expressed as a Chern-Simons term
\be\label{cs1}
\mathcal{H}=\frac1{32\pi^2}\int d^3u\epsilon_{\mu\nu\lambda}A_{\mu}F_{\nu\lambda}=1.
\ee

\section{topological soliton solution based on Hopf mapping}
\label{solu}

We consider a Yang-Mills-Higgs model in the $3+1$D spacetime with the lagrangian given as follows
\be
L=-\frac12D_{\mu}\phi^a\cdot (D_{\mu}\phi^a)^{\dagger}-\frac14F_{\mu\nu}^aF_{\mu\nu}^a
-V(\phi),
\ee
where $\phi^a$ is in the adjoint representation of a SU(2) gauge group such that $D_{\mu}\phi^a=\partial_{\mu}\phi^a+e\epsilon^{abc}A^b_{\mu}\phi^c$, and $V(\phi)=\lambda(\phi^a\phi^a-v^2)^2$.
We want to construct an instanton-type solution, the physical meaning of which will be clarified later. To achieve this, we consider the Euclidean version of the above model, which is equivalent to consider the spatial part of the 4+1D model. The spatial coordinates are chosen as $x_i$ for $i=1,2,3,4$. We introduce $z_1=x_1+ix_2$ and $z_2=x_3+ix_4$ and define $r^2=\sum_i|z_i|^2$. For simplicity, we set $v=1$ and $e=1$. As $r\to\infty$, $\phi^a$ approaches the classic vacuum solution as $\lim_{r\to\infty}\phi^a(\mathbf{x})=m^a(\mathbf{x})$ with $m^am^a=1$ so that the potential $V(\phi)$ is minimized.
Therefore, when $r\to\infty$, the vacuum solution is in fact a Hopf mapping which reads $S^3\overset{m^a}{\to} S^2$. We can define the map $m^a$ as $m^a(\mathbf{x})=\frac{\bar{z}_i\,\sigma^a_{ij}z_j}{r^2}$. Now the SU(2) gauge symmetry is spontaneously broken to the U(1) symmetry.

To minimize the total energy of the instanton, one need to require $D_{\mu}\phi^a=0$ in the limit of large $r$. Next we multiply both sides of $D_{\mu}\phi^a=0$ by $\epsilon^{ija}m^j$, and use the identity $\epsilon^{ija}\epsilon^{abc}=\delta^{ib}\delta^{jc}-\delta^{jb}\delta^{ic}$ to get
\be
A^i_{\mu}=-\epsilon^{ija}m^j\partial_{\mu}m^a+A^{b}_{\mu}m^bm^i.
\ee
If we can find a solution to $A^a_{\mu}$ which is perpendicular to $m^a$ in the space of vacua, i.e., $A^a_{\mu}m^a=0$, then we get a simple expression of $A^i_{\mu}=-\epsilon^{ija}m^j\partial_{\mu}m^a$, i.e., $A^a_{\mu}$ is a large gauge transformation as $r\to\infty$. For convenience, we define the shorthand notation that $\partial_{\mu}=\p_i,\bar{\p}_i$ with $\p_i=\frac{\p}{\p z_i}$ and $\bar{\p}_i=\frac{\p}{\p \bar{z}_i}$.
Hence the derivatives of $\phi^a$ at $r\to \infty$ are
\be \label{dm1}
\bar{\p}_k m^a=\frac{\sigma^a_{kj}z_j}{r^2}-m^a\frac{z_k}{r^2},\qquad
\p_k m^a=\frac{\sigma^a_{jk}\bar{z}_j}{r^2}-m^a\frac{\bar{z}_k}{r^2}
\label{dm}.
\ee
By using the identity (\ref{p2}), one can further verify that
\be
\bar{A}^a_k
&=&-i\frac{1}{r^4}(\sigma^a_{il}\delta_{jk}-\sigma^a_{kj}\delta_{il})\bar{z}_iz_jz_l\nonumber\\
&=&-i\frac{1}{r^2}(m^az_k-\sigma^a_{kj}z_j)\nonumber\\
&=&i\bar{\p}_km^a.
\ee
By a similar calculation, we also get $A^a_k=-\epsilon^{abc}m^b\p_km^c=-i\p_km^a$. Since $m^am^a=1$, then $m^a\p_{\mu}m^a=0$, therefore the solution does satisfy $A^a_{\mu}m^a=0$ and our derivation is indeed self-consistent. From above discussions we see that $\p_{\mu}\phi^a$ approaches 0 as fast as $1/r$ when $r\to \infty$ while $D_{\mu}\phi^a$ vanishes identically.

Now we know the needed asymptotic behaviors of the fields $\phi^a$ and $A^a_{\mu}$. To find the full instanton solution we adopt the ansatz
\be
&&\phi^a(\mathbf{x})=f(r)m^a(\mathbf{x}),\quad
A^a_k(\mathbf{x})=-i\p_km^a(\mathbf{x})g(r),\quad
\bar{A}^a_k(\mathbf{x})=i\bar{\p}_km^a(\mathbf{x})g(r),
\ee
where the continuous function $f$ and $g$ are required to satisfy $f(r)\to1$, $g(r)\to 1$ as $r\to\infty$ and $f(0)=g(0)=0$ (so that the scalar and gauge fields have well behaviors at the origin).

To evaluate the total action of the instanton, we need to know the various derivatives of $m^a$, which are outlined in Appendix.\ref{app1}. Using these results, it is straightforward to get the covariant derivatives
\be
\bar{D}_k\phi^a&=&
\bar{\p}_k m^a\,f+m^a\frac{z_k}{2r}f'+i\epsilon^{abc}\bar{\p}_km^bm^c\,gf\nonumber\\
&=&\bar{\p}_k m^a\,f(1-g)+m^a\frac{z_k}{2r}f',\nonumber\\
D_k\phi^a&=&\p_k m^a\,f(1-g)+m^a\frac{\bar{z}_k}{2r}f',
\label{dphi}
\ee
and the first term of the lagrangian is given by
\be
D_{\mu}\phi^a (D_{\mu}\phi^a)^{\dagger}=4\bar{D}_k\phi^a D_k\phi^a=\frac{8f^2(1-g)^2}{r^2}+(f')^2.
\ee
Using Eq.(\ref{t3}), the field strength is evaluated as
\be
F^a_{i\bar{j}}&=&\p_i\bar{A}^a_j-\bp_jA^a_i+\epsilon^{abc}A_i^b\bar{A}_j^c\nonumber\\
&=&2i\p_i\bp_jm^a\,g+\epsilon^{abc}\p_i m^b\bar{\p}_{j}m^c\,g^2
+\frac{i\bar{z}_i\bp_jm^a}{2r}g'+\frac{iz_j\p_im^a}{2r}g'\nonumber\\
&=&\frac{2i(r^2\delta_{ij}-\bar{z}_iz_j)}{r^4}m^a(g^2-2g)
+(\frac{i\bar{z}_i\bp_jm^a}{2r}+\frac{iz_j\p_im^a}{2r})g'.
\ee
To simplify our notation, we define the coefficient of $g'$ as
\be
C^a_{i\bar{j}}\equiv\frac{i\bar{z}_i\bp_jm^a}{2r}+\frac{iz_j\p_im^a}{2r}=i\frac{\bar{z}_i\sigma^a_{jk}z_k
+\bar{z}_k\sigma^a_{ki}z_j-2m^a\bar{z}_iz_j}{2r^3}.
\ee
It's easy to see that $C^a_{i\bar{i}}=0$ and $m^aC^a_{i\bar{j}}=0$. Similarly, by making use of Eq.(\ref{t4}), the other component of the field strength with two holomorphic or two anti-holomorphic indices are given by
\be
F^a_{ij}=-i(\frac{\bar{z}_i\p_jm^a}{2r}-\frac{\bar{z}_j\p_im^a}{2r})g', \quad F^a_{\bar{i}\bar{j}}=i(\frac{z_i\bp_jm^a}{2r}-\frac{z_j\bp_im^a}{2r})g'.
\ee
The expressions of all the field strength in the complex indices are list in Appendix.\ref{app1}. The squares of the field strength can be simplified by noting the following relations between the field strengths with complex and real indices
\be
&&F^a_{z_1,z_2}=\frac14(F^a_{13}-F^a_{24}-iF^a_{14}-iF^a_{23}),\nonumber\\
&&F^a_{z_1,\bar{z}_2}=\frac14(F^a_{13}+F^a_{24}+iF^a_{14}-iF^a_{23}),\nonumber\\
&&F^a_{z_1,\bar{z}_1}=\frac i2F^a_{12},\quad
F^a_{z_2,\bar{z}_2}=\frac i2F^a_{34}.
\ee
From these relations, we find
\be
\sum_{\mu,\nu}F^2_{\mu\nu}=8\sum_{i,j}
(F_{z_i,\bar{z}_j}F_{\bar{z}_i,z_j}+F_{\bar{z}_i,\bar{z}_j}F_{z_i,z_j}).
\ee
Since $m^aC^a_{i\bar{j}}=0$, the cross term vanishes in the square of $F^a_{i,\bar{j}}$. Using the fact that $z_i\p_im^a=\bar{z}_i\bar{\p}_im^a=0$, it is easy to find that
\be
\left|\frac{2i(r^2\delta_{ij}-\bar{z}_iz_j)}{r^4}\right|^2=\frac{4}{r^4},\quad
C^a_{\bar{i}j}C^a_{i\bar{j}}=\frac1{r^2}.
\ee
Thus we find
\be
F^a_{\bar{i}j}F^a_{i\bar{j}}=\frac{4(2g-g^2)^2}{r^4}+\frac{(g')^2}{r^2}.
\ee
SImilarly, we have
\be
F^a_{\bar{i}\bar{j}}F^a_{i,j}=\frac1{4r^2}(z_i\bp_jm^a-z_j\bp_im^a)
(\bar{z}_i\p_jm^a-\bar{z}_j\p_im^a)(g')^2=\frac{(g')^2}{r^2}.
\ee


Collecting all the above results, we finally get the total energy (or total action in the Euclidean space) as
\be
S=\int r^3dr\Big[\frac{4f^2(1-g)^2}{r^2}+\frac{(f')^2}{2}
+4\Big(\frac{2(2g-g^2)^2}{r^4}+\frac{(g')^2}{r^2}\Big)
+\lambda(f^2-1)^2\Big].
\ee
This result is quite similar to the magnetic monopole energy. The difference is that here we have to integrate the whole 3-sphere. Therefore the action is actually logarithm divergence. This situation is very similar to the vortex solution of scalar $O(2)$ model without coupling any gauge field in the two dimensional case. Therefore this type of soliton or instanton may not contribute in a infinitely large system. To get some physical effect, we should consider a finite sized system such that the radial integral has an upper bound. Then it makes sense to minimize the total action to find the instanton solutions.

\begin{figure}
\centerline{\includegraphics[width=0.5\textwidth]{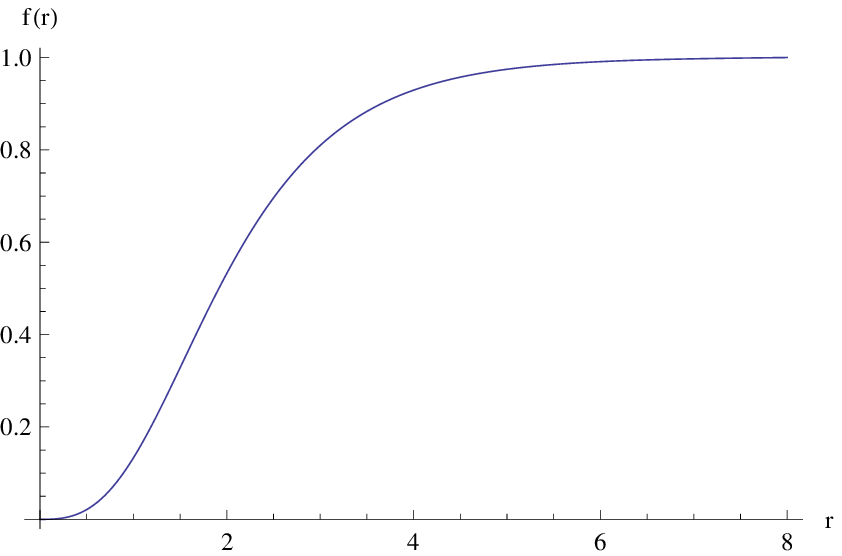}
\includegraphics[width=0.5\textwidth]{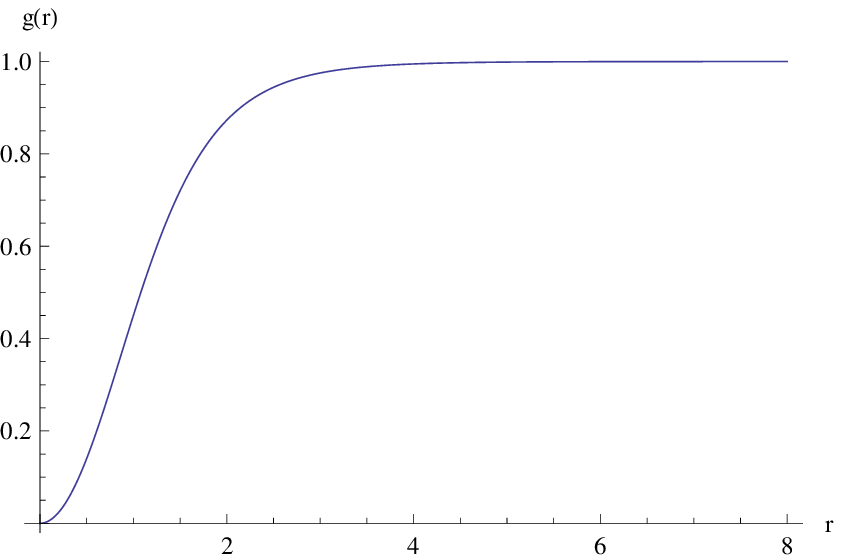}}
\caption{Numerical results of functions $f(r)$ and $g(r)$ v.s. $r$}
\label{fg}
\end{figure}
As usual, we use the variational method to find that $f,g$ satisfy the following
equations
\be
&&\frac{d^2g}{dr^2}+\frac1r\frac{dg}{dr}+f^2(1-g)-\frac{4(1-g)(2g-g^2)}{r^2}=0,\nonumber\\
&&\frac{d^2f}{dr^2}+\frac3r\frac{df}{dr}-\frac{8f(1-g)^2}{r^2}
-4\lambda f(f^2-1)=0.
\ee
with the boundary conditions $g(\infty)=f(\infty)=1$ and $f(0)=g(0)=0$.
These two equations are coupled nonlinear differential equations, which in general can not be solved analytically. Here we numerically solve the above differential equations. For convenience, we have taken $\lambda=1$. The numerical results are shown in Figure.\ref{fg}. Here we use a large fixed upper bound $r_c$ to replace the infinity.

\section{Topological charge and its physical meaning}
\label{topo}

For the t'Hooft-Polyakov magnetic monopole, the magnetic charge is also the winding number of the mapping that maps the spatial boundary onto the vacuum manifold. Hence the magnetic charge is the topological charge as well, thus it is quantized naturally. For our soliton-type solution, the topological charge is obviously related to the Hopf invariant. While its physical meaning is not as intuitive as that of the magnetic charge. This can be discussed by considering what happens at the boundary of the spacetime. The $SU(2)$ gauge symmetry breaks down into the $U(1)$ gauge symmetry there as pointed out in the previous section. For magnetic monopoles, it is this $U(1)$ gauge field that gives rise to a hedgehog-like magnetic field configuration. For our soliton-type solution, we identify what we introduced at the end of Section.\ref{hopfmap} to compute the Hopf invariant as this $U(1)$ gauge field .

We first give a warm-up discussion on the monopole. At the boundary, $A^a_{\mu}$ is designed to cancel $\p_{\mu}\phi^a$. The most possible form of $A^a_{\mu}$ that satisfies this requirement is
\be
A^a_{\mu}=\epsilon_{abc}\phi^b\p_{\mu}\phi^c+\phi^aA^{(1)}_{\mu}.
\ee
Here $A^{(1)}_{\mu}=A_{\mu}^a\phi^a$ is the gauge field associated with the unbroken U(1) symmetry. Then the corresponding field strength is
\be
F^a_{\mu\nu}=\phi^a\mathcal{F}_{\mu\nu},\quad
\mathcal{F}_{\mu\nu}=\p_{\mu}A^{(1)}_{\nu}-\p_{\nu}A^{(1)}_{\mu}
+\epsilon_{abc}\phi^a\p_{\mu}\phi^b\p_{\nu}\phi^c.
\label{Fmn}
\ee
However, the relation between the field strength and the associated vector potential is not simply $\mathcal{F}_{\mu\nu}=\p_{\mu}A^{(1)}_{\nu}-\p_{\nu}A^{(1)}_{\mu}$ as in the situation of the usual U(1) symmetry. The extra term $\epsilon_{abc}\phi^a\p_{\mu}\phi^b\p_{\nu}\phi^c$ in Eq.(\ref{Fmn}) cannot be written as the form $\p_{\mu}A_{\nu}-\p_{\nu}A_{\mu}$ with some vector potential $A_{\mu}$. In the mathematical language, this term is closed but not exact. It is precisely this term that is responsible for the monopole-like field configuration and makes the crucial contribution to the topological charge. The magnetic charge of the monopole is given by the integration of $\mathcal{F}_{\mu\nu}$ over the spatial boundary
\be
g=\int dS_{\mu\nu}\mathcal{F}_{\mu\nu}=\int dS_{\mu\nu}
\epsilon_{abc}\phi^a\p_{\mu}\phi^b\p_{\nu}\phi^c.
\ee
Clearly, the first term of $\mathcal{F}_{\mu\nu}$ does not make any contribution. The second term is the winding number of the vacuum configuration mapping as we mentioned before.

On the contrary, for our Hopf soliton solution, the extra term in Eq.(\ref{Fmn}) can be expressed as a curl of a vector potential due to the pulling back of the Hopf mapping, as we discussed in Section.\ref{hopfmap}.
Thus it can be expressed as
\be
\epsilon_{abc}\phi^a\p_{\mu}\phi^b\p_{\nu}\phi^c=\p_{\mu}A^{(2)}_{\nu}-\p_{\nu}A^{(2)}_{\mu}.
\ee
Then the field strength associated with the unbroken $U(1)$ symmetry is
\be
\mathcal{F}_{\mu\nu}=\p_{\mu}\mathcal{A}_{\nu}-\p_{\nu}\mathcal{A}_{\mu},\quad
\mathcal{A}_{\mu}=A^{(1)}_{\mu}+A^{(2)}_{\mu}.
\ee
In terms of $\mathcal{F}_{\mu\nu}$ and $\mathcal{A}_{\mu}$, we can construct the Hopf invariant
as in Section.\ref{hopfmap}
\be
\mathcal{H}=\frac1{32\pi^2}
\int d^3x\epsilon_{\mu\nu\lambda}\mathcal{A}_{\mu}\mathcal{F}_{\nu\lambda}.
\label{hopfnum}
\ee
Here the integral is over the three dimensional boundary of the 4D spacetime.
Since $A^{(1)}_{\mu}$ is topologically trivial, it is easy to see that terms like  $\epsilon_{\mu\nu\lambda}A^{(1)}_{\mu}\p_{\nu}A^{(1)}_{\lambda}$, $\epsilon_{\mu\nu\lambda}A^{(1)}_{\mu}\p_{\nu}A^{(2)}_{\lambda}$ and $\epsilon_{\mu\nu\lambda}A^{(2)}_{\mu}\p_{\nu}A^{(1)}_{\lambda}$ do not contribute to the above integral. By the construction of the Hopf soliton, we have
\be
\epsilon_{abc}\phi^a\p_{\mu}\phi^b\p_{\nu}\phi^c\Big|_{r\to \infty}
=\epsilon_{abc}m^a\p_{\mu}m^b\p_{\nu}m^c=\p_{\mu}A^{(2)}_{\nu}-\p_{\nu}A^{(2)}_{\mu}
\ee
with $A^{(2)}_{\mu}=-i\Big[\bar{\zeta_i}(\p_{\mu}\zeta_i)-(\p_{\mu}\bar{\zeta_i})\zeta_i\Big]$ where $\zeta_i=z_i/r$.
Therefore, we have
\be
\mathcal{H}=\frac1{16\pi^2}
\int d^3x\epsilon_{\mu\nu\lambda}A^{(2)}_{\mu}\p_{\nu}A^{(2)}_{\lambda}=1
\ee
according to Eq.(\ref{cs1}).
This result can also be directly obtained from $\mathcal{F}_{\mu\nu}$ at the boundary. All components of $\mathcal{F}_{\mu\nu}$ can be obtained from the expressions in Appendix.\ref{app3} by noting $g(\infty)=1$ and $g'(\infty)=0$
\be
&&\mathcal{F}_{12}=\frac{4(x_3^2+x_4^2)}{r^4},\quad
\mathcal{F}_{13}=\mathcal{F}_{24}=\frac{4(x_1x_4-x_2x_3)}{r^4},\nonumber\\
&&\mathcal{F}_{34}=\frac{4(x_1^2+x_2^2)}{r^4},\quad
\mathcal{F}_{14}=-\mathcal{F}_{23}=-\frac{4(x_1x_3+x_2x_4)}{r^4}.
\ee
If we express $\mathcal{F}_{\mu\nu}$ by a differential form, we find
\be
\mathcal{F}=\frac12\mathcal{F}_{\mu\nu}dx_{\mu}\wedge dx_{\nu}=-\omega_2.
\ee
Here $\omega_2$ is the pulled back volume element form as we defined in Section.\ref{hopfmap}. Follow the same steps, we find
\be
\mathcal{H}=\frac1{16\pi^2}\int \mathcal{A}\wedge\mathcal{F}
=\frac1{16\pi^2}\int \omega_1\wedge\omega_2=1.
\ee

To understand the physical meanings of this topological charge and the Hopf soliton solution, we must have a better understanding about the geometric meaning of the Hopf mapping. Since it is very difficult to visualize a 3-sphere embedded inside $\mathbb{R}^4$, we consider a deformed and simplified version of the Hopf mapping. We first deform $S^3$ into a cylinder $S^2\times [0,1]$, then treat $S^2$ as a one-point-compactification of $\mathbb{R}^2$. Now we show the following map: $\mathbb{R}^2\times [0,1] \to S^2$ has a nontrivial Hopf number
\be
&&y_1=\frac1r\sin[f(r)]\Big(x_1\cos[a(x_3)]-x_2\sin[a(x_3)]\Big),\nonumber\\
&&y_2=\frac1r\sin[f(r)]\Big(x_1\sin[a(x_3)]+x_2\cos[a(x_3)]\Big),\nonumber\\
&&y_3=\cos[f(r)].
\label{H1}
\ee
Here $(x_1,\,x_2)\in\mathbb{R}^2$, $x_3\in [0,1]$ and $r=\sqrt{x_1^2+x_2^2}$. We also assume that $a(t)$ is a monotonic function from $[0,1]$ to $[0,2\pi]$ and $f(0)=\pi$, $f(\infty)=0$.

This mapping is slightly different from the Hopf mapping. If we treat $x_3$ as a time variable, the above mapping describes that $\mathbb{R}^2$ makes a $2\pi$ rotation when time evolves from 0 to 1. Hence the world line of each point in $\mathbb{R}^2$ produces a helix curve. If we identify the initial time with the final time, the world line becomes a closed loop. Moreover, the world lines of two different points become two linked loops. In this sense, we expect that this map can give the similar result as the Hopf mapping.

The volume element or the field strength is determined by $F_{\mu\nu}=\epsilon^{ijk}y_i\p_{\mu}y_j\p_{\nu}y_k$. Thus we find
\be
F_{12}=\frac1r\sin[f(r)]f'(r),\quad F_{23}=\frac{x_2}r\sin[f(r)]f'(r)a'(x_3),\quad F_{31}=-\frac{x_1}r\sin[f(r)]f'(r)a'(x_3).
\ee
The corresponding vector potentials are
\be
A_1=\frac{x_2}{r}\cos[f(r)],\quad
A_2=-\frac{x_1}{r}\cos[f(r)],\quad
A_3=-a'(x_3)\cos[f(r)].
\ee
Finally we find
\be
\mathcal{H}=\frac1{16\pi^2}\int d^3x\epsilon_{\mu\nu\lambda}A_{\mu}F_{\nu\lambda}
=\frac1{8\pi^2}\int d^3x\frac{1}r\sin[f(r)]f'(r)a'(x_3)=1
\ee
This result reflects that the linking number of two world lines is 1 just as that of the Hopf mapping. Hence the map (\ref{H1}) is topologically equivalent to the Hopf mapping. This can also be understood more geometrically as we will state below.

We recall that $S^3$ can be decomposed into two solid tori. In complex coordinates, $S^3$ is described by $|z_1|^2+|z_2|^2=1$. Then the two solid tori are
\be
T_1: 1/2<|z_1|^2<1,\quad |z_2|^2=1-|z_1|^2;\\
T_2: 0<|z_1|^2<1/2,\quad |z_2|^2=1-|z_1|^2.
\ee
There are two types of nontrivial cycles on the torus which are also the generators of the $\pi_1(T^2)$. These two tori $T_{1}$ and $T_2$ are related by a modular transformation which exchanges the two types of cycles. If the torus is characterized by a complex number $\tau$, then this modular transformation is given by $\tau\to-\frac{1}{\tau}$. It is easy to verify that, under the Hopf mapping, the pre-image of the northern hemisphere $S^N$ is just $T_1$ and that of the southern hemisphere $S^S$ is $T_2$.

It can be found that the pre-image of a fixed point on $S^N$ is a circle described by $(z_1e^{i\phi},z_2e^{i\phi})$. Here $e^{i\phi}$ is an arbitrary phase factor and $z_{1,2}$ are fixed complex numbers solved from the Hopf mapping equations. If we trace the trajectory of the vector $(z_1e^{i\phi},z_2e^{i\phi})$, we find that it makes a $2\pi$ rotation on the $z_1$ plane, and a $2\pi$ rotation on the $z_2$ plane simultaneously. The resulting curve is a helix with the starting and ending points identified. If we cut the torus into a cylinder, then we retrieve the helix world line. Since a solid torus can be treated as $D^2\times S^1$ ($D^2$ is a 2D disc), each point on $S^N$, under the map $D^2\times S^1\to S^N$, corresponds to a point on $D^2$ which makes a $2\pi$ rotation as we travel along $S^1$.
This is also true for the southern part $S^S$. After making a modular transformation, we can glue the southern part back to the northern part to get a map like $S^2\times S^1\to S^2$. If we cut $S^1$ into a interval $[0,1]$, we retrieve the map $S^2\times [0,1]\to S^2$. Now it is easy to see that each point on $S^2$ makes a $2\pi$ rotation as we travel along the interval $[0,1]$. Therefore this map is indeed described by Eq.(\ref{H1}). Hence the geometric meaning of the Hopf mapping is clearly understood.

Comparing with the fact that the $S^2\to S^2$ map describes the magnetic monopole, the Hopf soliton associated with the nontrivial $S^3\to S^2$ map describes the tunneling event that the monopole makes a $2\pi$ rotation of the vacuum manifold $S^2$. Since in the monopole solution, the spatial boundary and the vacuum manifold are locked together, then this event is equivalent to make a spatial $2\pi$ rotation.

If there is no Hopf number dependent term in the Hamiltonian, this tunneling event still has no direct physical effects. We know that the Hopf number term (\ref{hopfnum}) is expressed as the surface integral of the boundary of the 4D spacetime. One might guess that the corresponding term in the bulk will be the $\theta$ vacuum term
\be
\mathcal{H}=\frac{\theta}{32\pi^2}
\int d^4x\epsilon_{\mu\nu\lambda\rho}F^a_{\mu\nu}F^a_{\lambda\rho},
\ee
where $\theta$ is an arbitrary angle, since $\mathcal{F}$ is the unbroken component of the $F^a$ at the spacetime boundary. However, a direct evaluation shows that the term $F^a\wedge F^a$ vanishes identically. This means that $\mathcal{F}$ cannot be simply replaced by $F^a$. Therefore, the correct term is
\be
\mathcal{H}=\frac{\theta}{32\pi^2}\int d^4x\epsilon_{\mu\nu\lambda\rho}\mathcal{F}_{\mu\nu}\mathcal{F}_{\lambda\rho}
=\frac{\theta}{32\pi^2}\int d^4x\epsilon_{\mu\nu\lambda\rho}
\mathcal{F}^a_{\mu\nu}\mathcal{F}^b_{\lambda\rho}\phi^a\phi^b.
\ee
This term is a higher order term which usually does not appear in ordinary gauge theories. But it may possibly appear in some low energy effective theory. If such terms appear, the monopole will pick up a nontrivial phase factor when making a $2\pi$ rotation, which changes the statistics of monopoles to that of anyons.

\section{Conclusion and discussion}
\label{conclu}
We have constructed the Hopf soliton solution in the 3+1D non-Abelian gauge theory based on a nontrivial Hopf mapping. The topological charge is identified with the Chern-Simons term of the unbroken $U(1)$ gauge field, which corresponds to some higher order term of the non-Abelian gauge field. The Hopf soliton is a spacetime event that executes a $2\pi$ rotation of the monopole. The appearing of the Hopf soliton and Hopf term together may generate the fractional statistics for the monopole. It seems contradict the spin-statistics theorem, but it is possible that finite-sized objects can acquire some exotic statistics law in a finite system.

Since monopoles has not been discovered in nature yet, all the above discussions may seem to be of purely academic interests.
However, non-Abelian monopoles can be realized in certain condensed matter systems such as superflud $A$ phase of $^3$He. Thus, in a finite system of $^3He$\cite{Zhang}, it is quite possible that Hopf soliton may have real physical effects if the Hopf term appears in the low energy theory.

We would like to thank Chih-Chun Chien for useful discussion. Hao Guo thanks the support by NSF of China (Grants No. 11204032, SBK201241926) and by the Fundamental Research Funds for the Central Universities.

\appendix
\section{Some useful formulas}\label{app1}
In the main text, we make frequent use of the following identities of Pauli matrices
\be
&&\sigma^a_{ij}\sigma^a_{kl}=2\delta_{il}\delta_{jk}-\delta_{ij}\delta_{kl}.\label{p1}\\
&&\epsilon^{abc}\sigma^b_{ij}\sigma^c_{kl}=i(\sigma^a_{il}\delta_{jk}-\sigma^a_{kj}\delta_{il}). \label{p2}
\ee

In computing the covariant derivative and field strength, we need to calculate the 2nd order derivatives of $m^a$.
When analyzing the asymptotic behavior of $A^a_{\mu}$, we have got the first order derivatives of $m^a$
\be\label{appe1}
\p_im^a=-i\epsilon^{abc}m^b\p_im^c,\quad \bar{\p}_im^a=i\epsilon^{abc}m^b\bar{\p}_im^c.
\ee
From the identity $m^a\p_{\mu}m^a=0$, we can directly find the following results
\be \label{appe2}
z_i=m^a\sigma^a_{ij}z_j,\quad \bar{z}_i=\bar{z}_jm^a\sigma^a_{ji}.
\ee
When evaluating the term $D_{\mu}\phi^a(D_{\mu}\phi^a)^{\dagger}$, we need to calculate the product of two first order derivatives of $m^a$. By using Eqs.(\ref{dm1}), (\ref{p1}) and (\ref{appe2}), we have
\be\label{t0}
\p_im^a\p_jm^a&=&\frac{1}{r^4}\Big(\sigma^a_{pi}\sigma^a_{qj}\bar{z}_p\bar{z}_q-m^a\sigma^a_{pi}\bar{z}_p\bar{z}_j-m^a\sigma^a_{qj}\bar{z}_i\bar{z}_q+\bar{z}_i\bar{z}_j\Big)\nonumber\\
&=&\frac{1}{r^4}\Big[(2\delta_{pj}\delta_{iq}-\delta_{pi}\delta_{qj})\bar{z}_p\bar{z}_q-\bar{z}_i\bar{z}_j\Big]\nonumber\\
&=&0.
\ee
Similarly, we can get $\bar{\p}_im^a\bar{\p}_jm^a=0$. These results obviously imply $\p_im^a\p_im^a=\bar{\p}_im^a\bar{\p}_im^a=0$ immediately. Another type of product is evaluated similarly as
\be
\p_im^a\bar{\p}_jm^a&=&\frac{1}{r^4}\Big[(2\delta_{pq}\delta_{ij}-\delta_{pi}\delta_{jq})\bar{z}_pz_q-\bar{z}_iz_j\Big]\nonumber\\
&=&\frac{2(r^2\delta_{ij}-\bar{z}_iz_j)}{r^4}.
\ee
This immediately implies $\p_im^a\bar{\p}_im^a=\frac{2}{r^2}$.

When computing the term $F^a_{\mu\nu}F^a_{\mu\nu}$, we must know the quantities like $\p_{\mu}\p_{\nu}m^a$ and $\epsilon^{abc}m^a\p_{\mu}m^b\p_{\nu}m^c$. We don't need to worry about $\p_i\p_jm^a$ since they are cancelled in the expressions of $F^a_{ij}$. To evaluate $\p_i\bar{\p}_jm^a$, we first start from the expression (\ref{dm1}) and take one more derivative
\be
\p_i\bar{\p}_jm^a=\frac{\sigma^a_{ji}}{r^2}-\frac{\sigma^a_{jk}z_k\bar{z}_i}{r^4}-\frac{\sigma^a_{ki}z_j\bar{z}_k}{r^4}+2m^a\frac{\bar{z}_iz_j}{r^4}.
\ee
We further have
\be \label{t1}
\epsilon^{abc}m^b\p_i\bar{\p}_jm^c
&=&\epsilon^{abc}\Big(\frac{\sigma^b_{pq}\sigma^c_{ji}\bar{z}_pz_q}{r^4}
-\frac{\sigma^b_{pq}\sigma^c_{jk}\bar{z}_pz_qz_k\bar{z}_i}{r^6}
-\frac{\sigma^b_{pq}\sigma^c_{ki}\bar{z}_pz_qz_j\bar{z}_k}{r^6}+2m^bm^c\frac{\bar{z}_iz_j}{r^4}\Big)\nonumber\\
&=&0,
\ee
where the properties $\epsilon^{abc}m^bm^c=0$, $\bar{z}_iz_i=r^2$ and Eq.(\ref{p2}) have been applied. We can also start from the second identity of Eqs.(\ref{appe1}) to calculate the second order derivative of $m^a$
\be\label{t3}
\p_i\bar{\p}_jm^a&=&i\epsilon^{abc}\p_im^b\bar{\p}_jm^c\nonumber\\
&=&\epsilon^{abc}\epsilon^{bpq}m^p\p_im^q\bar{\p}_jm^c=(\delta^{aq}\delta^{cp}-\delta^{ap}\delta^{cq})m^p\p_im^q\bar{\p}_jm^c\nonumber\\
&=&-\frac{2(r^2\delta_{ij}-\bar{z}_iz_j)}{r^4}m^a,
\ee
where Eq.(\ref{t1}) has been applied in the second line, $m^a\p_{\mu}m^a=0$ has been applied in the last line. To determine $\epsilon^{abc}m^a\p_im^b\p_jm^c$, we use Eq.(\ref{appe1}) again
\be \label{t4}
\epsilon^{abc}m^a\p_im^b\p_jm^c&=&-im^a\epsilon^{abc}\epsilon^{bpq}m^p\p_im^q\p_jm^c\nonumber\\
&=&im^a(m^a\p_im^c\p_jm^c-m^c\p_im^a\p_jm^c)\nonumber\\
&=&0,
\ee
where the equality (\ref{t0}) has been applied. Similarly, we can show that $\epsilon^{abc}m^a\bar{\p}_im^b\bar{\p}_jm^c=0$.

For completeness, we list all the components of field strength as follows
\be
F^a_{12}&=&\frac{\bar{z}_i(\sigma^2\sigma^a)_{ij}\bar{z}_j}{2r^3}g',\\
F^a_{\bar{1}\bar{2}}&=&\frac{z_i(\sigma^a\sigma^2)_{ij}z_j}{2r^3}g',\\
F^a_{1\bar{1}}&=&\frac{2i|z_2|^2m^a\,(2g-g^2)}{r^4}
+i\frac{r^4\delta_{a3}-(\bar{z}_i\sigma^a_{ij}z_j)(\bar{z}_i\sigma^3_{ij}z_j)}{2r^5}g',\\
F^a_{1\bar{2}}&=&\frac{-2i\bar{z}_1z_2m^a\,(2g-g^2)}{r^4}
+i\frac{r^4\delta^a-2\bar{z}_1z_2(\bar{z}_i\sigma^a_{ij}z_j)}{r^5}g',\\
F^a_{2\bar{1}}&=&\frac{-2i\bar{z}_2z_1m^a\,(2g-g^2)}{r^4}
+i\frac{r^4\bar{\delta}^a-2\bar{z}_2z_1(\bar{z}_i\sigma^a_{ij}z_j)}{r^5}g',\\
F^a_{2\bar{2}}&=&\frac{2i|z_1|^2m^a\,(2g-g^2)}{r^4}
-i\frac{r^4\delta_{a3}-(\bar{z}_i\sigma^a_{ij}z_j)(\bar{z}_i\sigma^3_{ij}z_j)}{2r^5}g'.
\ee
Here we define $\delta^a=(1,\,i,\,0)$ and $\bar{\delta}^a=(1,\,-i,\,0)$.

\end{document}